# Reliability

This special volume of *Statistical Sciences* presents some innovative, if not provocative, ideas in the area of reliability, or perhaps more appropriately named, integrated system assessment. In this age of exponential growth in science, engineering and technology, the capability to evaluate the performance, reliability and safety of complex systems presents new challenges. Today's methodology must respond to the ever-increasing demands for such evaluations to provide key information for decision and policy makers at all levels of government and industry—problems ranging from international security to space exploration. We, the co-editors of this volume and the authors, believe that scientific progress in reliability assessment requires the development of processes, methods and tools that combine diverse information types (e.g., experiments, computer simulations, expert knowledge) from diverse sources (e.g., scientists, engineers, business developers, technology integrators, decision makers) to assess quantitative performance metrics that can aid decision making under uncertainty. These are highly interdisciplinary problems. The principal role of statistical sciences is to bring statistical rigor, thinking and methodology to these problems.

Bedford, Quigley and Walls open the issue by reviewing the role of expert judgment to support reliability assessments within the systems engineering design process, with an aim toward developing a framework for tracking reliability through the design process. Chambers, James, Lambert and Vander Wiel develop methods for a contemporary software issue—monitoring the health of networked applications. Peña reviews recent work in the modeling and analysis of recurrent events, and Lee and Whitmore review first hitting time models, with special emphasis on regression structures. Both of these papers have wide relevance in diverse fields, including engineering, reliability, biomedicine and public health. Wilson, Graves, Hamada and Reese review methodology for addressing system reliability with limited full system testing, including resource allocation considerations. Lindqvist reviews basic modeling approaches for failure and maintenance data from repairable systems, with emphasis on imperfect repair models and the trend-renewal process. Escobar and Meeker review accelerated testing methods, which help estimate the failure-time distributions or long-term performance of components of high-reliability products.

This volume presents some forward looking ideas from an international set of researchers that are focused on the growing challenges facing statistical sciences in the domain of integrated system assessment. We hope you enjoy the material.

<div style="text-align:right">
Sallie Keller-McNulty, Co-Editor<br>
Alyson Wilson, Co-Editor<br>
Christine Anderson-Cook, Co-Editor
</div>

*Sallie Keller-McNulty is Dean of the George R. Brown School of Engineering, Rice University, Houston, Texas 77005–1892, USA (e-mail: sallie@rice.edu). Alyson Wilson is a Technical Staff Member, Statistical Sciences Group, MS F600, Los Alamos National Laboratory, Los Alamos, New Mexico 87545, USA (e-mail: agw@lanl.gov). Christine Anderson-Cook is a Technical Staff Member, Statistical Sciences Group, MS F600, Los Alamos National Laboratory, Los Alamos, New Mexico 87545, USA (e-mail: c-and-cook@lanl.gov).*